\documentclass[12pt]{iopart}

\usepackage{latexsym}
\usepackage{graphics}
\usepackage{graphicx}
\usepackage{amssymb}
\usepackage{dcolumn}
\usepackage{bm}
\usepackage[francais]{babel}
\usepackage{psfrag}
\usepackage{hyperref}
\usepackage{color}
\usepackage[latin1]{inputenc}

\begin{document}

\title{Attractive and repulsive cracks in a heterogeneous material}
\author{Pierre-Philippe Cortet, Guillaume Huillard, Lo\"{i}c Vanel and Sergio Ciliberto}

\address{Laboratoire de physique, CNRS UMR 5672, Ecole Normale
Sup\'erieure de Lyon, Universit\'e de Lyon, 46 all\'ee d'Italie,
69364 Lyon Cedex 07, France\ead{Loic.Vanel@ens-lyon.fr}}

\date{Received: date / Revised version: date}

\begin{abstract}
We study experimentally the paths of an assembly of cracks growing
in interaction in a heterogeneous two-dimensional elastic brittle
material submitted to uniaxial stress. For a given initial crack
assembly geometry, we observe two types of crack path. The first
one corresponds to a repulsion followed by an attraction on one
end of the crack and a tip to tip attraction on the other end. The
second one corresponds to a pure attraction. Only one of the crack
path type is observed in a given sample. Thus, selection  between
the two types appears as a statistical collective process.
\end{abstract}


\maketitle

\section{Introduction}

Understanding the rupture mechanisms of solids has become an
important goal of fracture physics in order to improve the
strength of structures and avoid catastrophic failures.
Characterization of rupture properties very often involve the
growth of a dominant crack. For instance, there is an extensive
literature, both experimental and theoretical, discussing the slow
growth dynamics of a single crack along a straight path in brittle
\cite{Hsieh,Rice,Marder,Santucci1,Santucci2,Santucci2b,Santucci2c}
or visco-plastic materials
\cite{Schapery,Schapery2,Schapery3,Kaminskii,Kaminskii2,Chudnovsky,Cortet1,Cortet2}.
However, in many practical situations, the crack path is not
straight. For one thing, the crack path can be slightly
destabilized and develop some roughness due to dynamical
instabilities \cite{Boudet} or material heterogeneities. This kind
of path instability has motivated a large amount of experimental
works analyzing the roughness of cracks
\cite{Schmittbuhl,Boffa,Salminen,Mallick,Santucci4,Celarie,Bonamy}
as well as several models describing roughening mechanisms
\cite{Ramanathan,Schmittbuhl2,Bonamy2} and non-trivial effects of
heterogeneities on the rate of crack growth
\cite{Kierfeld,Cortet3,Guarino}. Much larger deviations of the
crack path from a straight line can be observed during the growth
of an array of interacting cracks. Comparatively to the case of
roughness, this rather complex situation has been studied
essentially theoretically
\cite{Melin,Hori,Kachanov,Kachanov2,Seelig,Xia,Brener} and very
little experimentally \cite{Eremenko,Dolotova,Broberg} despite its
practical importance especially for heterogeneous materials where
multiple cracks are likely to form. Understanding the growth of
interacting cracks is also a relevant issue for fault dynamics
\cite{An} as well as crack pattern formation during drying
processes \cite{Groisman,Shorlin,Bohn,Bohn2}.

There are several levels of complexity that can play a role in the
growth of interacting cracks. First, depending on the geometry of
the crack array and the loading, the stress field around a crack
will be amplified or shielded because of the existence of other
cracks. Then, in $2d$, the effective contribution of the loading
on each crack will usually be a mixture of mode I and mode II
\footnote{The mode I corresponds to a tensile stress normal to the
plane of the crack as the mode II corresponds to a shear stress
acting parallel to the plane of the crack and perpendicular to the
crack front.}. It has been argued that crack deviations occur when
the shear stress on the crack lips (mode II) is non-zero and that
the crack will grow so as to minimize the mode II contribution and
maximize the mode I contribution \cite{Erdogan,Goldstein,Rice2}. A
noticeable observation related to this property is the
preferential merging of cracks joining each other forming a right
angle, commonly observed for crack patterns in drying experiments
\cite{Shorlin,Bohn,Bohn2}. This occurs naturally because the
principal stress along a crack lip is parallel to the crack
direction so that a crack approaching the lip with a right angle
is propagating mainly in mode I. On the other hand, when two
collinear mode I cracks are growing towards each other, they do
not merge tip to tip, but instead repel each other
\cite{Eremenko}. The origin of this effect has been discussed by
several authors \cite{Melin,Kachanov} and numerical simulations
have been able to reproduce, at least qualitatively, experimental
observations \cite{Seelig}. However, there are not many
experimental data to compare with these theoretical predictions,
and furthermore, there is very little knowledge about the effect
of material heterogeneities on the crack path selection.

In this paper, we study experimentally the trajectories of
interacting cracks in an almost two dimensional brittle and
heterogeneous material. The array of cracks has been initiated
prior to the application of the external stress and the
heterogeneity of the material (paper sheets) allows us to test the
stability of crack paths to small perturbations. We uncover that
for a given geometry of the crack array, a crack can follow
statistically two stable trajectories : an attractive one and
a repulsive one with respect to the neighboring cracks. The main
result of this investigation is the analysis of the geometrical
conditions for which cracks are attracted towards another and when
they are repelled. The paper is organized as follows. In the next
section, we describe the experimental apparatus. In section 3,
the extraction procedure of the post-mortem shapes of the
crack profiles is described. In sections 4, 5 and 6, we analyze
the different types of crack paths and their statistics as a
function of the initial crack pattern geometry. In section 7, we
finally discuss the results and conclude.

\section{The experiments}

\begin{figure}[h!]
\psfrag{S}{$\sigma$} \psfrag{d}[l][][0.9]{$d$}
        \centerline{\includegraphics[width=7.5cm]{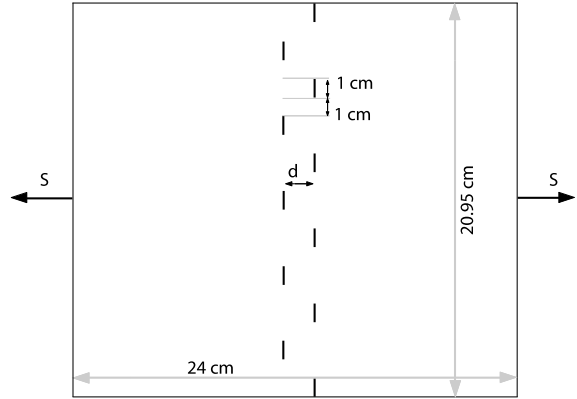}}
        \caption{Geometry of the crack array initiated in the
        paper sample and the direction of the applied stress $\sigma$.}
        \label{reseau}
\end{figure}

To study the growth of several interacting cracks, we have
loaded bi-dimensional brittle samples made of fax paper sheets
(Alrey\textsuperscript{\textregistered}) with a tensile machine.
The samples were previously prepared cutting several straight
cracks in the paper sheets. In order to observe clearly the
interaction between cracks during their growth, we had to find a
geometry for which all crack tips are equivalent. Actually, we
wanted to prevent the isolated growth at a particular crack tip
that will inhibit the growth of the other cracks. One possible
geometry that follows this condition consists in an array of
cracks as presented in figure \ref{reseau}. The pattern formed by
the two lines of cracks offers the advantage to have some
translational invariance in the cracks direction. The stress is
applied perpendicularly to the cracks direction and uniformly on
the sample borders. Therefore, stress intensification is
theoretically equivalent at each crack tip. In this geometry,
there are three adjustable parameters: the crack length, the
vertical spacing of cracks on a line and the horizontal spacing
between the two lines. In this article, the first two parameters
are fixed to 1cm and only the distance $d$ between the two lines
of cracks has been varied.

Each sample is prepared with the 11 initial cracks forming the
array using a cutter blade. The experimental set-up consists of a
tensile machine driven by a motor (Micro Controle UE42) controlled
electronically to move step by step (Micro Controle ITL09). The
paper sheets are mounted on the tensile machine with both ends
attached with glue tape and rolled twice over rigid bars clamped
on jaws. The motor controls the displacement of one jaw (400 steps
per micrometer) while the other jaw is rigidly fixed to a force
gage (Hydrotonics-TC). The tensile machine is placed in a box with
a humidity level stabilized at $5\%$ and at room temperature. The
samples are loaded by increasing the distance between the jaws on
which they are clamped such that the resulting force $F=\sigma e
H$ ($e=50\mu$m is the film thickness and $H=20.95$cm the sample
height) is perpendicular to the initial cracks direction. The
loading is enforced at a constant and very slow velocity (between
$0.84\mu$m.s$^{-1}$ and $2\mu$m.s$^{-1}$) until the final
breakdown of the sample. More details about the experimental
set-up can be found in \cite{Santucci2c}. Actually, once the force
threshold $F_c$ of the sample is overcome, we observe a
quasi-instantaneous rupture of the sample with a
quasi-simultaneous growth of all the cracks. There is no visual
evidence of any sub-critical crack growth process before the final
breakdown of the sample. During this breakdown, the applied force
drops. When it reaches about half its maximum
value, we stop loading the sample before breaking it in two
pieces. It is a convenient procedure that allows us afterwards to
image properly the crack path configurations in each sample.

\section{Post-mortem analysis of crack trajectories}

\begin{figure}[h!]
    \centerline{\includegraphics[width=12cm]{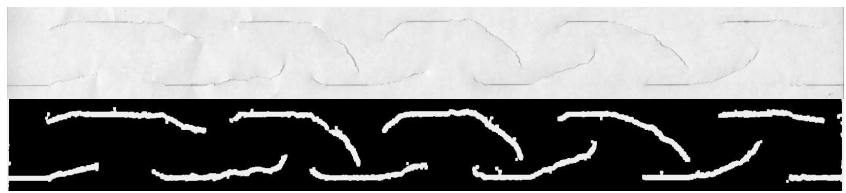}}
        \caption{Post-mortem images of a fractured sample with $d=1.5$cm:
        at the top, initial digitized image and, at the bottom,
        corresponding binary image with extracted crack paths obtained through image processing.}
        \label{analyseimg}
\end{figure}

\begin{figure}[h!]
\psfrag{Y}[l][][0.9]{$y$ (cm)} \psfrag{X}[l][][0.9]{$x$ (cm)}
        \centerline{\includegraphics[width=5.5cm]{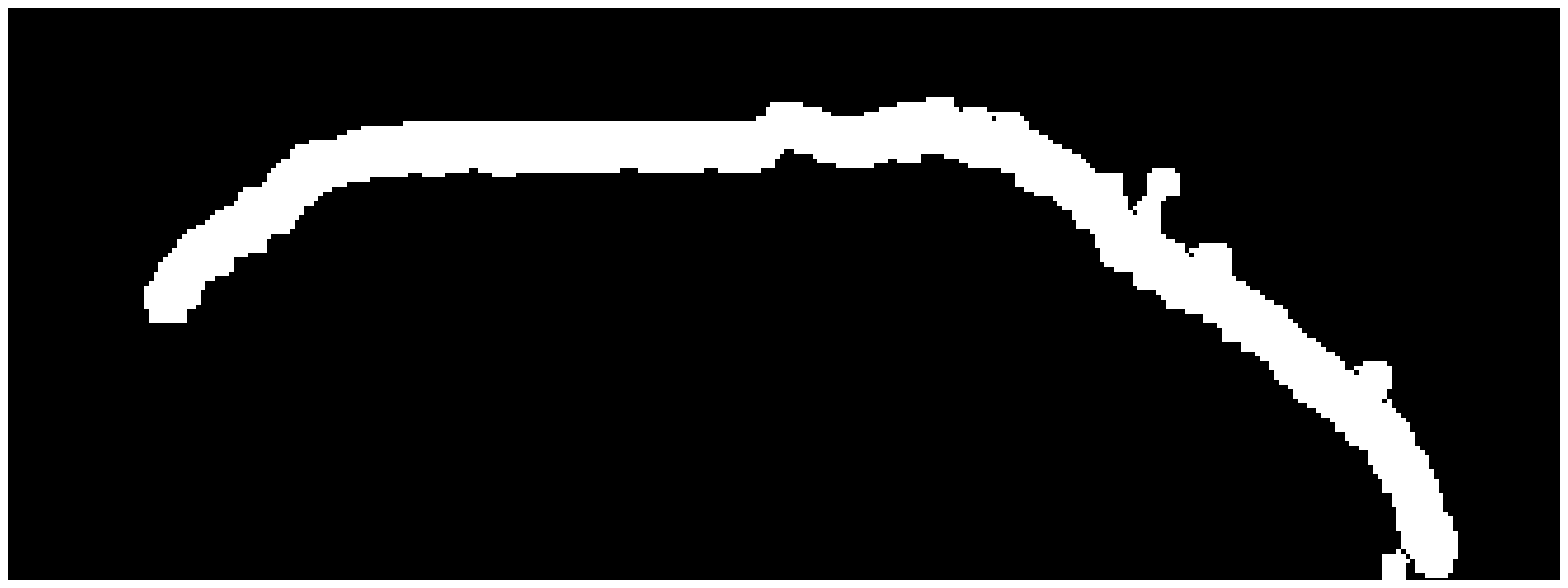}\hspace{0.5cm}
        \includegraphics[width=7.37cm]{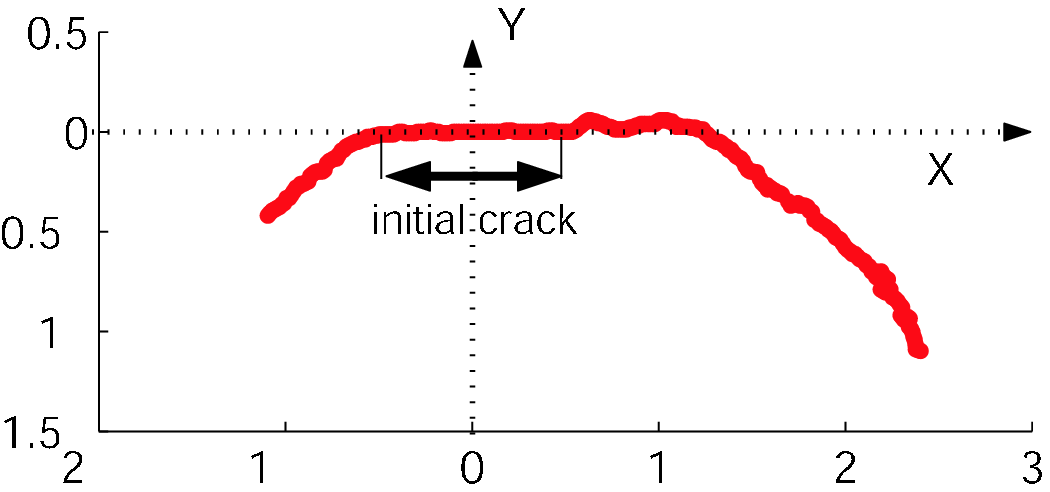}}
        \caption{\small On the left, binary image showing the path followed by a
        crack, and on the right, corresponding extracted crack profile $y(x)$.}
       \label{fonction}
\end{figure}

After each experiment, post-mortem samples are digitized using a
high resolution scanner. The obtained images are processed in
order to get a binary image in which the crack paths can be
distinguished. In figure \ref{analyseimg}, we show an example of a
digitized sample and of the corresponding extracted binary image.
This image processing allows to finally create for each crack path
the profile $y(x)$ that describes its shape as a function of the
abscissae $x$ along the axis corresponding to the initial crack
direction and centered on the corresponding initial crack (cf.
figure \ref{fonction}).

\section{Two types of crack path}

\begin{figure}[h!]
        \centerline{\includegraphics[width=12cm]{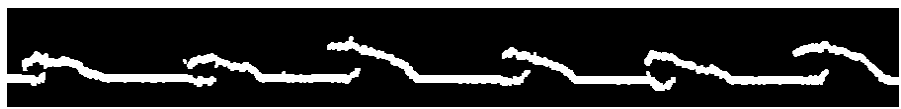}}
        \caption{Binary image of a fractured sample with $d=0$cm.}
       \label{repulsionbis}
\end{figure}

\begin{figure}[h!]
        \centerline{\includegraphics[width=3cm]{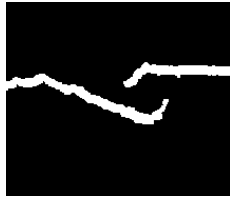}}
        \caption{Close-up of a binary image of a fractured sample with $d=0$cm.}
       \label{spirale}
\end{figure}

For an horizontal spacing $d=0$cm between the two crack lines, the
initial crack array collapses into a single crack line. In figure
\ref{repulsionbis}, one can see that the crack paths show a
repulsion phase before an attraction one. This initial crack
pattern always leads to the so-called type B$_0$ growth behavior.
This is the signature of the fact crack tips repulse each
other. Indeed, the crack tips move away from the initial crack
line before the neighboring tips overtake each other. Then, the
paths curve to join the neighboring crack lips tending to form a
right angle. Couples of crack paths appear to describe a sort of
spiral shape as one can see in figure \ref{spirale}.

\begin{figure}
     \centerline{\includegraphics[width=12cm]{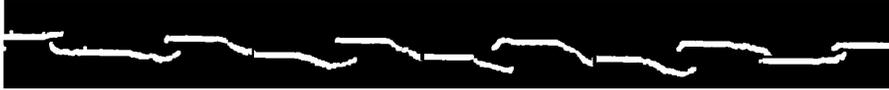}}
     \caption{Binary image of a fractured sample with $d=0.4$cm.}
       \label{repulsion0}
\end{figure}

For $0<d<d_{\rm{\tiny int}}$ (experimentally $d_{\rm{\tiny
int}}\approx 1.8 \pm 0.2$cm), two different kinds of behavior have
been identified with respect to the crack path shapes among the
fifty crack arrays that have been fractured. These two behaviors
occur statistically for a given sample geometry. For the first
one, that will be referred to as type A, neighboring cracks on
different lines attract each other from the beginning of the crack
growth up to the end as it is the case in figure \ref{analyseimg}
($d=1.5$cm). In contrast, for the type B, the crack paths are a
mixture of two paths, a repulsive plus attractive path on one end
and an attractive path tip to tip on the other end of each crack.
For instance, in figure \ref{repulsion0} ($d=0.4$cm), we observe a
repulsion phase for the right tip of the bottom cracks followed by
an attraction phase by the neighboring crack on the other line
while the right tip of the top cracks tends to join the left tip
of the bottom cracks in a rather unexpected way. It seems that the
symmetry breaking associated to the repulsive phase of crack
growth on one end of the crack allows simultaneously the merging
of tips on the other end of the crack.

\begin{figure}
        \centerline{\includegraphics[width=12cm]{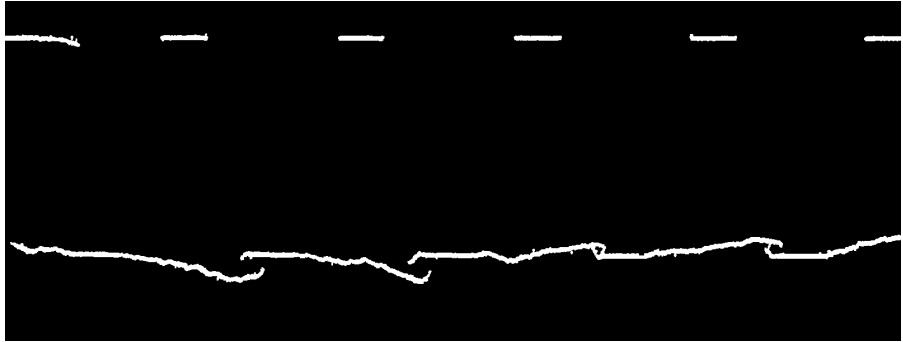}}
        \caption{\small Binary image of a fractured sample with $d=5$cm.}
       \label{indpt}
\end{figure}

Finally, when $d>d_{\rm{\tiny int}}$, the two lines of cracks do
not interact (cf. figure \ref{indpt}). Actually, one of the
two lines fractures preferentially and we get the type B$_0$ behavior
previously observed for $d=0$cm since the situation is comparable.

\begin{figure}
\centerline{\includegraphics[width=8.5cm]{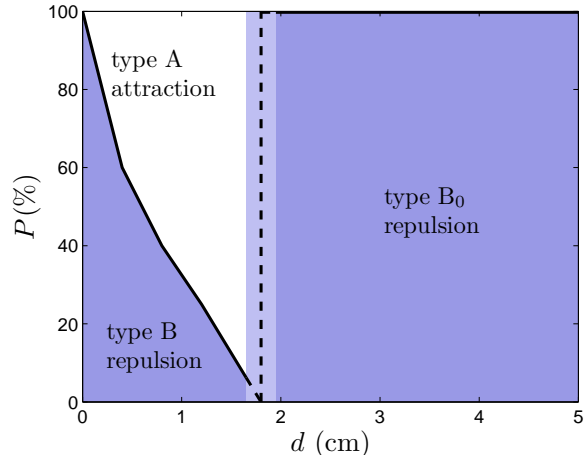}}
        \caption{Percentage of samples for which the crack path are of type B (including the case B$_0$). Experimentally, there
        is some uncertainty concerning the distance at which there is no more interactions
        between the two crack lines. This is schematically shown by the dotted line and lighter
        color band around $d=1.8$cm.}
       \label{pourcentage}
\end{figure}

To summarize the results:
\begin{itemize}
\item For $d=0$, only type B$_0$ is observed.
\item  For $0<d<d_{\rm{\tiny int}}$, types A and B are observed.
\item For $d_{\rm{\tiny int}}<d$, only type B$_0$ is observed.
\end{itemize}
The statistical proportion between types A and B is a function of
the spacing $d$ between the two initial crack lines. In figure
\ref{pourcentage}, we plot for each value of $d$ the percentage
$P$ of samples for which the fracture process follows type B (including the case B$_0$). This
percentage decreases regularly starting from 100
$\%$ for $d=0$ down to zero for $d=d_{\rm{\tiny int}}$ where it
jumps back brutally to $100\%$ (this last case corresponds to
rupture on a single crack line as observed in figure \ref{indpt}).

In the case of a homogeneous material, we would have expected the
crack trajectories to be reproducible from one experiment to the
other. A sheet of paper is actually made of a complex network of
cellulose fibers. Scanning electron microscopy on our samples
shows fiber diameters between 4 and 50$\mu$m with an average of
18$\mu$m. Cellulose fibers are themselves a bundle of many
microfibrils that have a crystalline structure \cite{Jakob}. The
statistical selection between two types of crack path shapes for a
given initial crack geometry is probably triggered by this
heterogeneity of the fractured material. Indeed, heterogeneity and
anisotropy in the initial crack tip toughness and shape due to the
heterogeneity of the material might induce small perturbations in
the initiation of the crack growth path. We discover that these
perturbations are large enough for the crack to explore
statistically two very different ``meta-stable'' crack
paths.

Additionally, the crack patterns observed in this section allow
us to confirm the fact that, in general, two crack tips
repulse each other while a crack tip and a crack lip attract each
other.

\section{Analysis of attractive cracks in type A samples}

\begin{figure}
\psfrag{Y}[c][][0.9]{$\langle y(x)\rangle$ (cm)}
\psfrag{X}[c][][0.9]{$x$ (cm)} \psfrag{Z}[c][][0.9]{$\langle
\langle y(x)\rangle \rangle$ (cm)} \psfrag{A}[c][][0.9]{(a)}
\psfrag{B}[c][][0.9]{(b)}
        \centerline{\includegraphics[width=7.3cm]{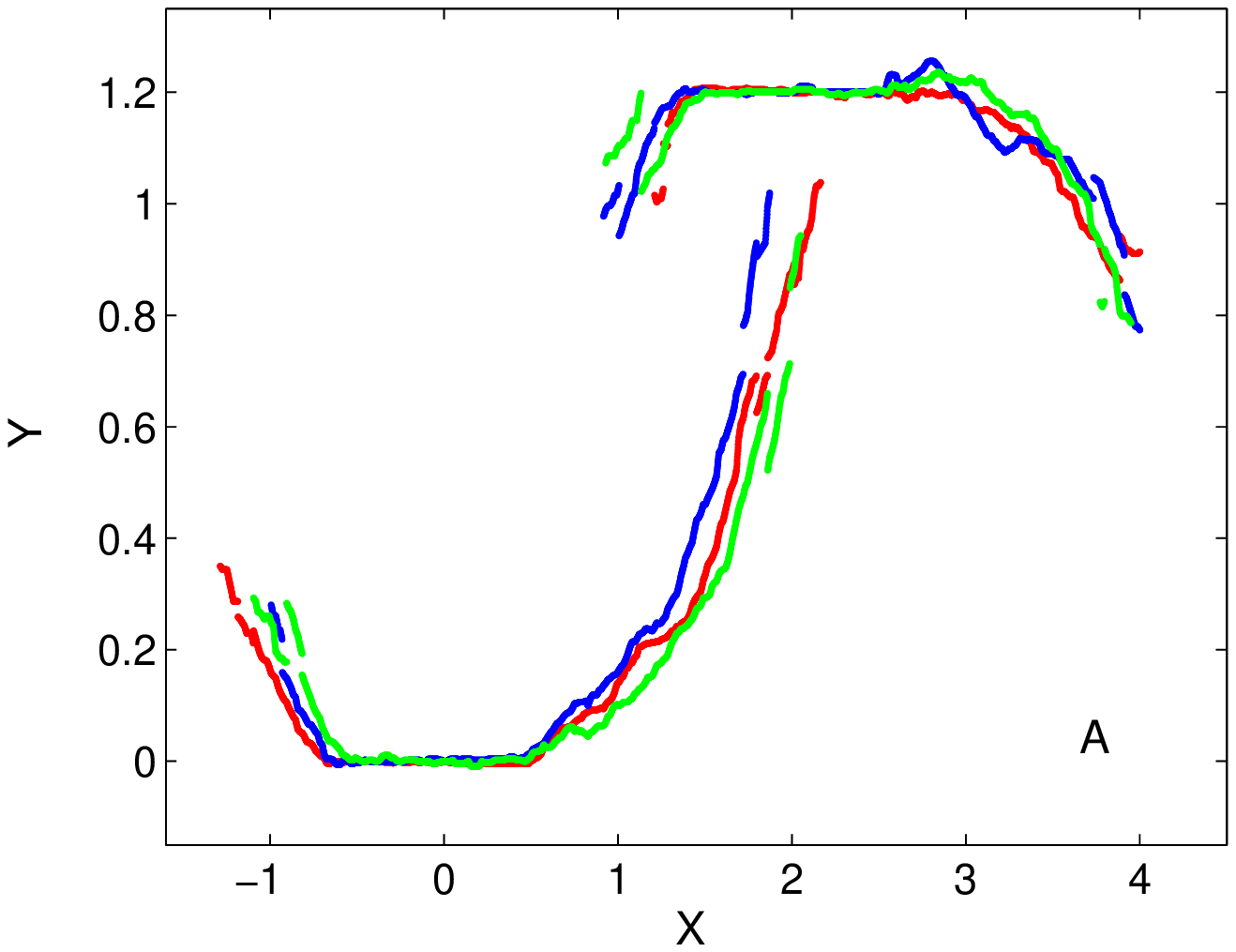}
        \hspace{0.1cm}
        \includegraphics[width=7.45cm]{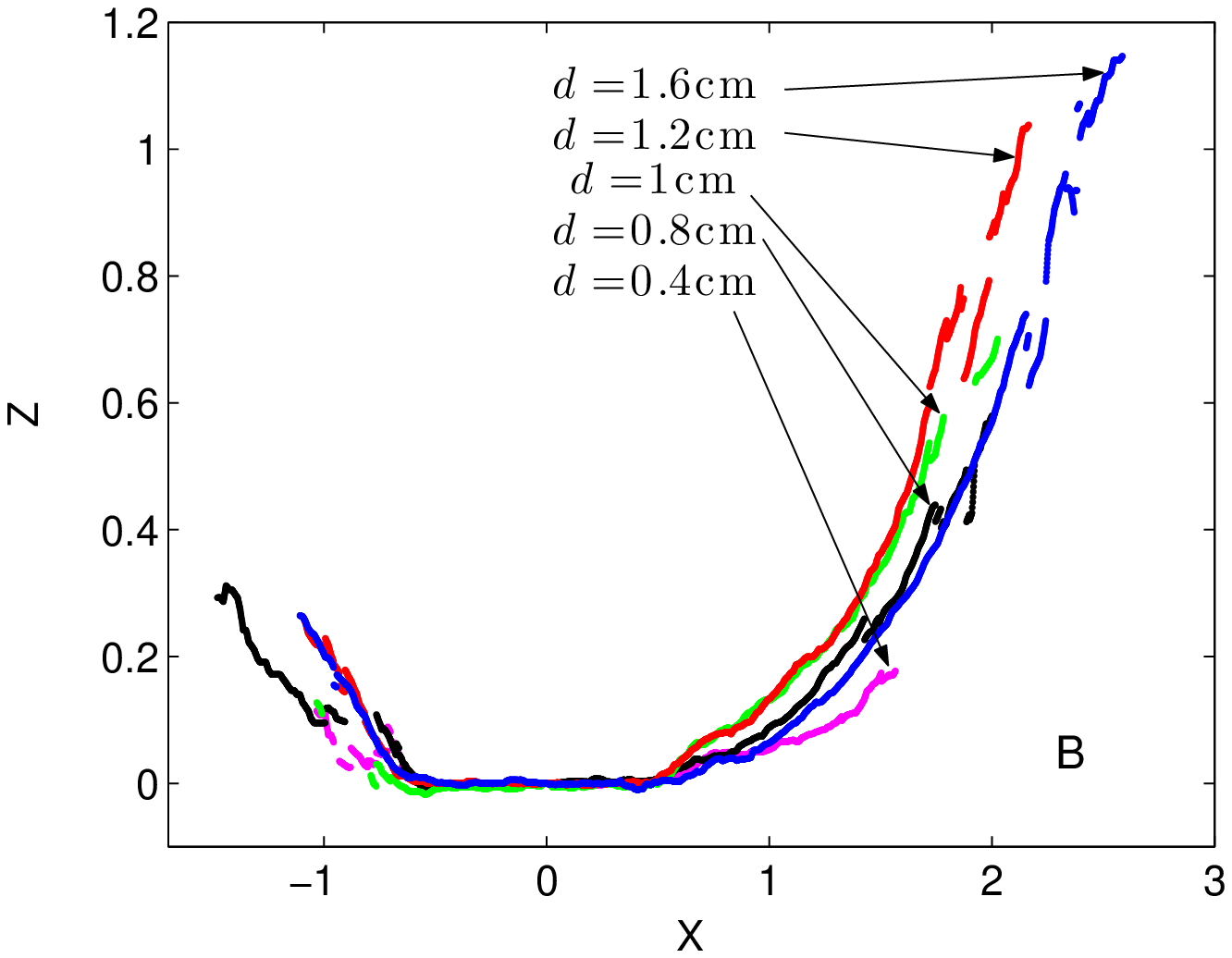}}
        \caption{(a) Average crack profile $\langle y(x)\rangle$
        for three samples with $d=1.2$cm and (b) average crack profile
        $\langle \langle y(x)\rangle \rangle$ for five values of $d$,
        for experiments with a type A behavior.}
         \label{attirance}
\end{figure}

In this section, we study the average behavior followed by type A
cracks that corresponds to experiments in which the cracks have
been attracting each other during their whole growth. For each
value of the spacing $d$ between the two crack lines, we extract
the average profile of the cracks. Actually, for each sample, we
compute the mean crack profile $\langle y(x)\rangle$ averaging
over the profiles $y(x)$ of all the cracks located on the same
initial crack line. In figure \ref{attirance}(a), we plot for three
different samples the averaged profiles of the two crack lines
separated by $d=1.2$cm. Then, we average the profiles $\langle
y(x)\rangle$ over all the type A samples corresponding to the same
value of $d$ (cf. figure \ref{attirance}(b)). We get the mean type A
profile $\langle \langle y(x)\rangle \rangle$ for each value of
$d$.

\begin{figure}
\psfrag{Y}[c][][0.9]{$\langle y(x)\rangle$ (cm)}
\psfrag{X}[c][][0.9]{$x$ (cm)} \psfrag{A}[c][][0.9]{(a)}
\psfrag{B}[c][][0.9]{(b)}\psfrag{L}[c][][0.9]{$\ell_j$
(cm)}\psfrag{d}[c][][0.9]{$d$ (cm)} \psfrag{U}[c][][0.9]{$\ell_j$}
        \centerline{\includegraphics[width=7.35cm]{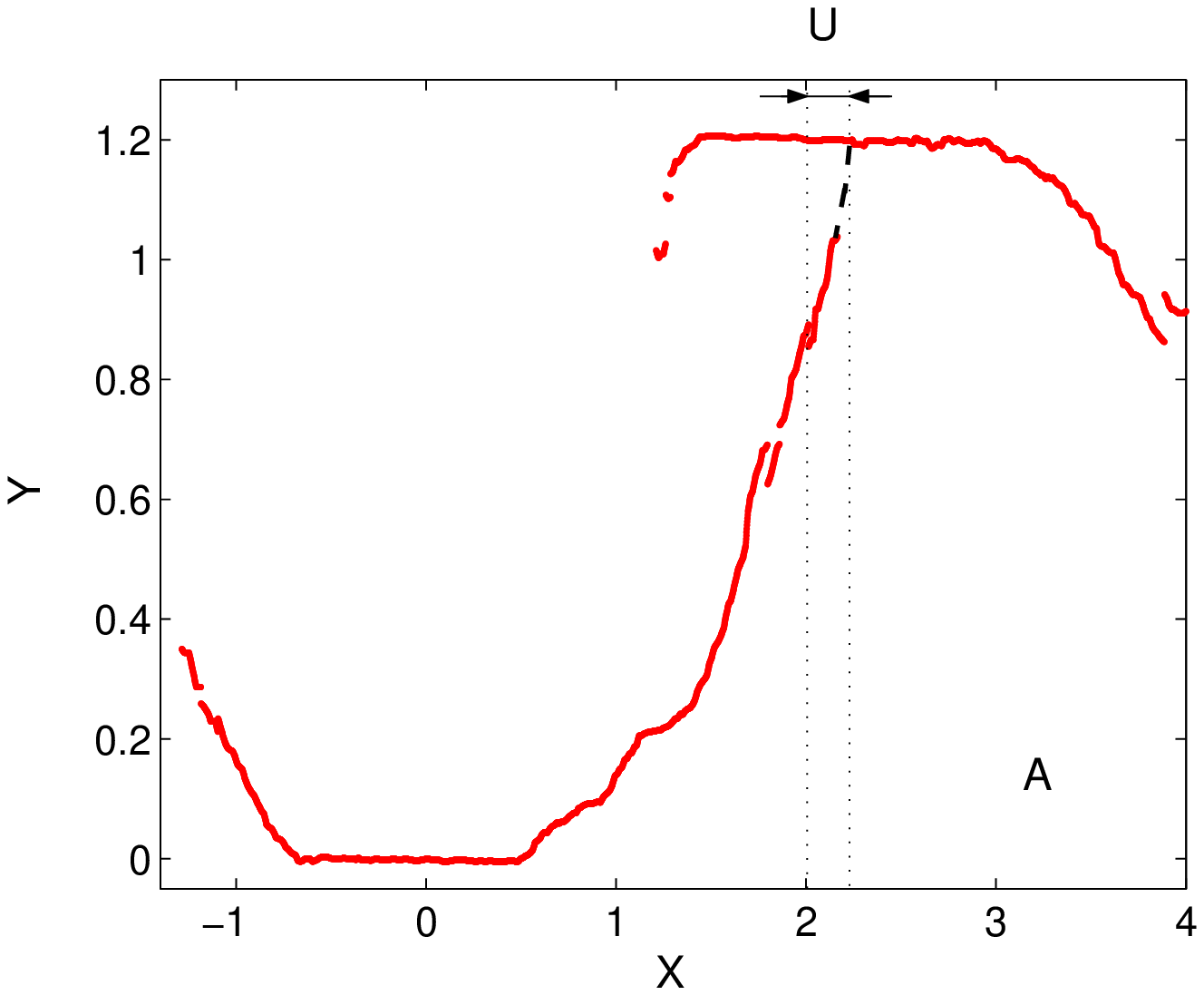}
        \hspace{0.1cm}
        \includegraphics[width=7.3cm]{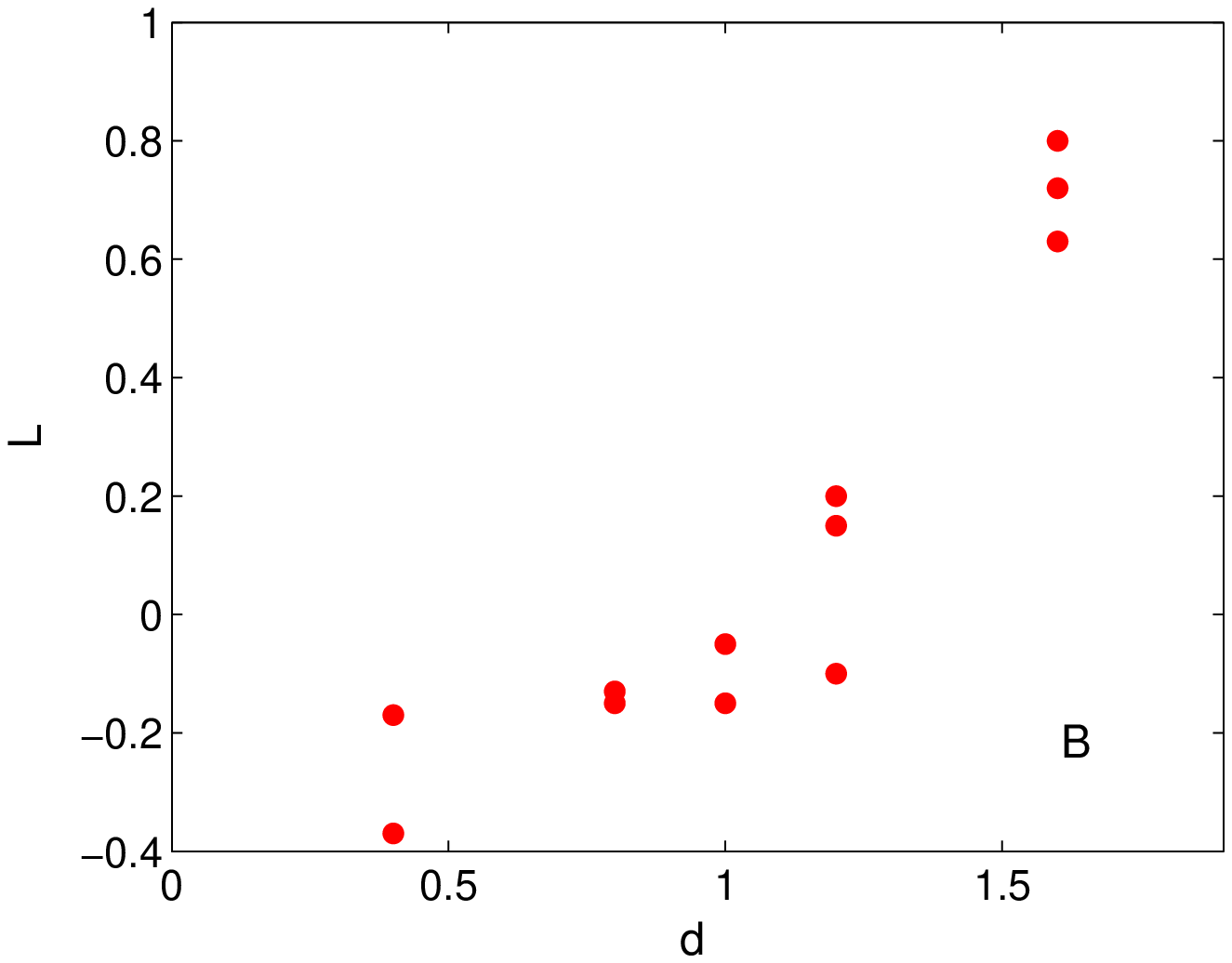}}
        \caption{(a) Average crack profile $\langle y(x)\rangle$
        for a sample with $d=1.2$cm and its extension allowing to define
        $\ell_j$ and (b) distance $\ell_j$ as a function of the spacing
         $d$ between initial crack
        lines, for experiments with a type A behavior. Each point corresponds to one experiment.}
         \label{attirance2}
\end{figure}

It is important to notice that, depending on the position $x$, the
number of profiles on which the averaging is performed is variable
since all the cracks do not have the same length in the $x$
direction. This is the reason why the averaged profiles, $\langle
y(x)\rangle$ and $\langle \langle y(x)\rangle \rangle$, are not
continuous at all points. For the same value of $d$, cracks of
different samples appear to a have very similar mean profiles
$\langle y(x)\rangle$ as one can see in figure \ref{attirance}(a).
Additionally, for all type A cracks, the dependence of the crack
path on the distance between the initial crack lines $d$ is weak
and non-systematic. The main difference is that the crack paths
extend farther away when the distance $d$ is increased (cf. figure
\ref{attirance}(b)).

Extending intuitively each crack path to join the neighboring
crack lip on the other crack line, we are able to define an
hypothetical junction position $\ell_j$ measured from the center
of the joined crack (cf. figure \ref{attirance2}(a)). In figure
\ref{attirance2}(b), we see that this junction distance $\ell_j$
increases rapidly with $d$. Actually, the rapid increase of this
distance between $d=1.2$cm and $d=1.6$cm suggests that there might
be a divergence of the junction position $\ell_j$ when the
distance $d$ approaches $d_{\rm{\tiny int}}(\approx 1.8$cm). A
divergence of $\ell_j$ would make sense since when $d>d_{\rm{\tiny
int}}$ the two crack lines do not interact anymore and thus, the
cracks of one line cannot cross those of the other line.

\section{Analysis of repulsive cracks in type B samples}

\begin{figure}
\psfrag{Y}[c][][0.9]{$\langle y(x)\rangle$ (cm)}
\psfrag{X}[c][][0.9]{$x$ (cm)} \psfrag{Z}[c][][0.9]{$\langle
\langle y(x)\rangle \rangle$ (cm)} \psfrag{A}[c][][0.9]{(a)}
\psfrag{B}[c][][0.9]{(b)}
        \centerline{\includegraphics[width=7.35cm]{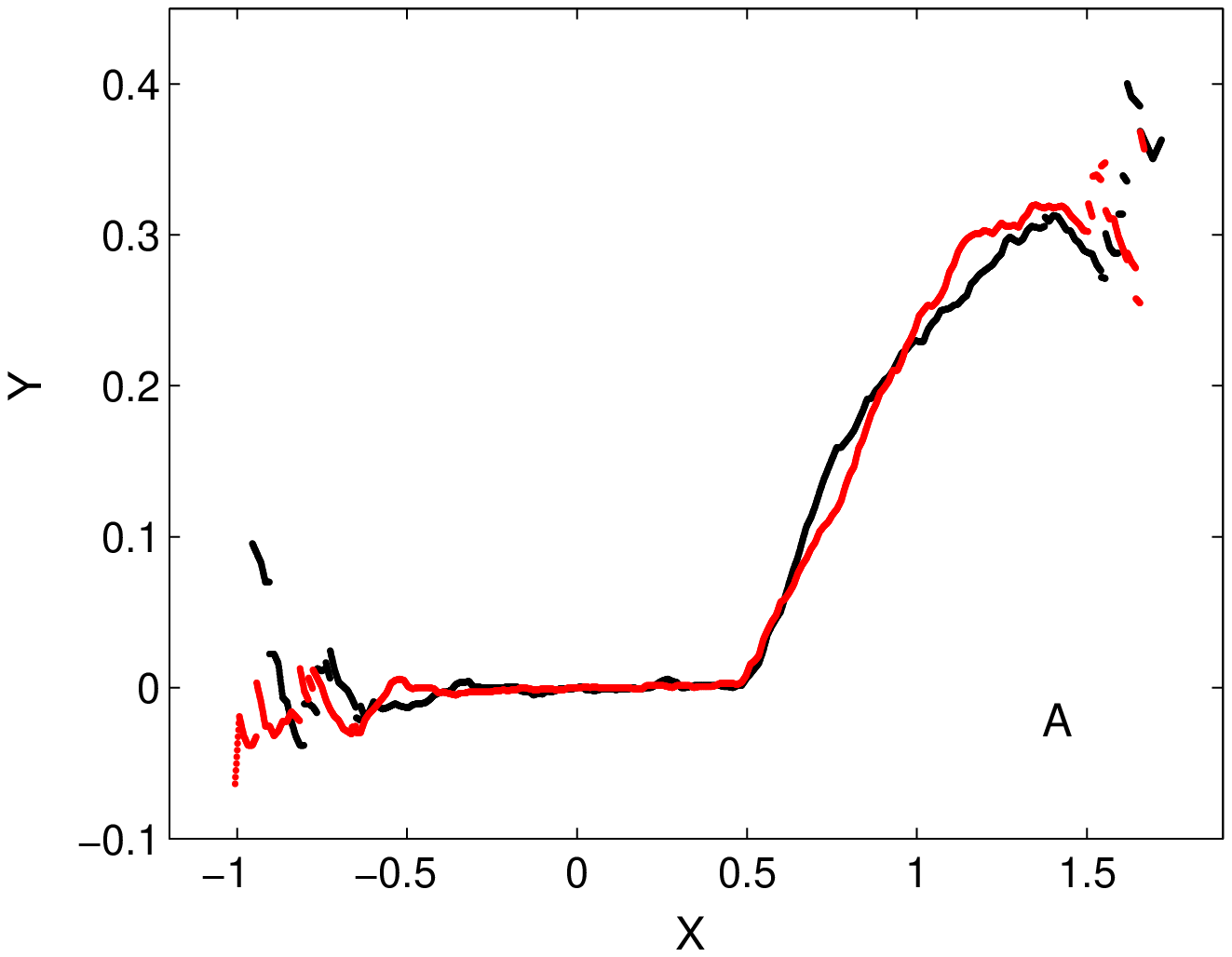}
        \hspace{0.1cm}
        \includegraphics[width=7.45cm]{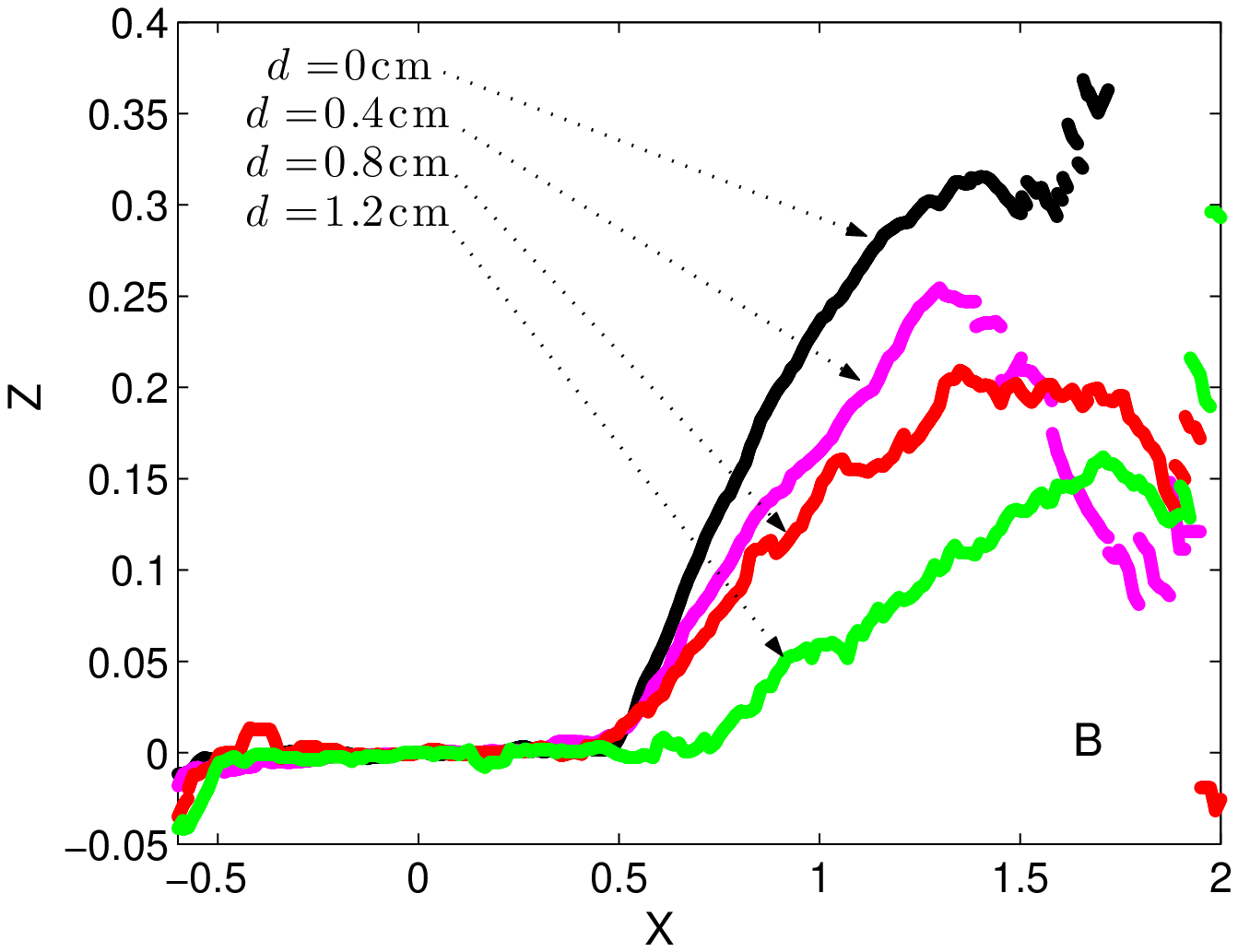}}
        \caption{Paths of repelled cracks (type B). (a) Average crack profile $\langle y(x)\rangle$
        for two samples with $d=1$cm, and (b) average crack profile
        $\langle \langle y(x)\rangle \rangle$ for four value of $d$.}
         \label{repulsion}
\end{figure}

\begin{figure}
        \centerline{\includegraphics[width=8cm]{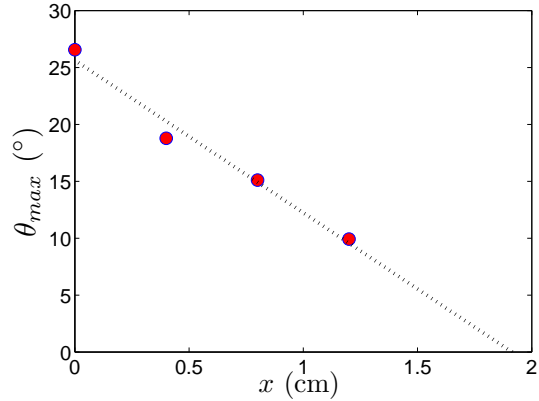}}
        \caption{Maximum angle of repulsion of the crack path
        during the repulsion phase as a function of the distance
        $d$ for type B experiments.}
         \label{etude}
\end{figure}

In this section, we study in type B (including B$_0$ for $d=0$)
samples the average behavior followed by the repulsive crack
paths. In figure \ref{repulsion}(a), we plot the average profile
of repulsive cracks $\langle y(x)\rangle$ for two samples with
$d=1$cm. We note that, for a given distance $d$, the crack path is
reproducible from one sample to the other. In figure
\ref{repulsion}(b), we plot the average crack profile $\langle
\langle y(x)\rangle\rangle$ as a function of $d$ and $x$. Clearly,
the smaller $d$, the larger the repulsion. One way to quantify
this is to measure the maximum angle of deviation
$\theta_{\tiny{max}}$ during the repulsion phase. As can be seen
in figure \ref{etude}, this quantity decreases with the distance
$d$. If we extend linearly this curve, we can see that the effects
of the repulsion disappear when $d$ reaches a value of about
$1.9$cm. Interestingly, we recover a value in the range
corresponding to the characteristic distance $d_{\rm{\tiny int}}$
above which the two crack lines do not interact.

For $d=0$, we find a maximum deviation angle $\theta_{\tiny{max}}\simeq
27^\circ$. This value is significantly larger than Kachanov's
prediction $\theta_{\tiny{max}}\simeq 14^\circ$ \cite{Kachanov}
that was obtained assuming the crack goes in the direction for
which the strain energy release rate would be maximum. Melin
\cite{Melin} has a different approach to explain the deviation of
collinear cracks. She analyzes the stability of the crack path to
a local perturbation. The crack deviation angle is then $ \theta =
\rm{atan} (2 \delta y/L)$ where $\delta y$ is the amplitude of the
local deviation and $L$ is the crack length. To obtain
$\theta_{\tiny{max}}\simeq 27^\circ$ would require a perturbation
$\delta y \simeq L/4 = 0.25$cm, a rather unphysical value. A large
deviation angle could also be obtained assuming that several
smaller local deviations of the crack path occur due to the
heterogeneous structure of paper. However, if that was the case,
one would expect the crack to progressively rotate instead of
turning rather abruptly (see figure \ref{repulsion}(b)).

\section{Discussion and conclusion}

In this article, we have seen that the interactions between two
lines of cracks growing under a very low strain rate in a
heterogeneous material lead to a statistical behavior. Indeed, for
the same initial sample geometry, we get two types of crack path
pattern. The type A behavior, that corresponds to a permanent
attraction between the cracks during their growth, appears to
present a universal shape that does not depend significantly on
the initial crack array geometry. On the contrary, for type B
behavior, for which the crack paths present either a repulsion
phase followed by an attraction or an attraction phase tip to tip,
the path trajectories appear to be dependent on the distance
between the two initial crack lines $d$. Indeed, the repulsion
decreases as $d$ increases and finally totaly disappears as $d$
tends to a certain distance $d_{\rm{\tiny int}}$. This particular
value of $d$ is also a critical value above which the two lines do
not interact anymore. Understanding this value is still an open
issue. Also, the value of the crack deviation angle observed in
our experiments during the repulsive phase remains unexplained. In
particular, it is rather large compared to theoretical
predictions.

The statistics observed for the crack path shape is the signature
of the meta-stability of two types of crack trajectory for a
given geometry. This bi-stability is for the moment an unexplained
property but it seems to be intrinsic to the chosen crack pattern
geometry. It is likely that the trigger for this
statistics comes from the complex structure of paper sheets that
makes this material heterogeneous and leads to anisotropy on the
initial crack tip shape and toughness. The two types of crack path
are two meta-stable paths between which each crack has to choose
when it initiates its growth. Only at this initiation stage, may
the heterogeneity of the sample strongly influence the crack in
its ``choice''.

It is also important to point out the collective behavior of the
cracks located on the same initial crack line. Indeed, inside a
given crack line, all the cracks select a specific dynamics with
quasi-identical crack path shape. Actually, the initiation of the
global rupture of a sample is probably triggered at one particular
crack tip. The choice for the crack shape type is made at this
particular moment. Then, an unexplained collective effect operates
to make all the cracks throughout a line behave the same way.

\ack{This work was funded with grant ANR-05-JCJC-0121-01.}

\end{document}